\documentclass{easychair}

\usepackage{latexsym}
\usepackage{amsfonts}
\usepackage{url}
\usepackage{xspace}
\usepackage{color}

\newif\ifdraft
\drafttrue           
\draftfalse

\newcommand{\VS}{\vspace{0pt}}
  \let\olditemize=\itemize
  \renewcommand{\itemize}{\olditemize\setlength{\itemsep}{2pt}}
  \renewcommand{\VS}{\vspace{2pt}}

\title{
  Coalescing: Syntactic Abstraction for Reasoning in First-Order Modal Logics
  \thanks{This work has been partially funded by the Microsoft Research-Inria Joint
    Centre, France. It has also been supported by the European Union Seventh
    Framework Programme under grant agreement no. 295261 (MEALS) and
    by the French BGLE Project ADN4SE.}
}
\titlerunning{Coalescing for Reasoning in First-Order Modal Logics}

\author{
  Damien Doligez\inst{1} \and
  Jael Kriener\inst{2} \and
  Leslie Lamport\inst{3} \and\\
  Tomer Libal\inst{2} \and
  Stephan Merz\inst{4}
}

\authorrunning{Doligez, Kriener, Lamport, Libal, Merz}

\institute{
  Inria, Paris, France \and
  MSR-Inria Joint Centre, Saclay, France \and
  Microsoft Research, Mountain View, CA, U.S.A. \and
  Inria, Villers-l\`es-Nancy, France
}

\renewcommand{\vec}[1]{\mathbf{#1}}

\renewcommand{\qed}{\hspace*{\fill}\textsc{q.e.d.}}
\renewcommand{\th}{\textsuperscript{th}\xspace}
\newcommand{\eqdef}{\ =_\textsf{\scriptsize\upshape def}\ }
\newcommand{\eps}{\epsilon}
\renewcommand{\implies}{\Rightarrow}
\newcommand{\tlaplus}{\mbox{TLA\kern -.35ex$^+$}\xspace}
\newcommand{\kw}[1]{\textsc{#1}}  
\newcommand{\ps}[2]{\ensuremath{\langle #1 \rangle #2}}
\newcommand{\nat}{\mathbb{N}}
\newcommand{\sem}[1]{\ensuremath{[\![ #1 ]\!]}}
\newcommand{\fun}{\rightarrow}
\newcommand{\true}{\textsf{tt}}
\newcommand{\false}{\textsf{ff}}
\newcommand{\FOL}[1]{\ensuremath{#1_{\textit{\scriptsize FOL}}}}

\newcommand{\ML}[1]{\ensuremath{#1_{\textit{\scriptsize ML}}}}

\newcommand{\folmodels}{\mathop{\models_{\textit{\scriptsize FOL}}}}

\newcommand{\nfolmodels}{\mathop{\not\models_{\textit{\scriptsize FOL}}}}
\newcommand{\mlmodels}{\mathop{\models_{\textit{\scriptsize ML}}}}

\newcommand{\nmlmodels}{\mathop{\not\models_{\textit{\scriptsize ML}}}}

\newcommand{\modal}{\nabla}
\newcommand{\dual}{\Delta}

\newcommand{\HH}{\mathcal{H}}
\newcommand{\II}{\mathcal{I}}
\newcommand{\KK}{\mathcal{K}}

\newcommand{\MM}{\mathcal{M}}

\newcommand{\OO}{\mathcal{O}}

\renewcommand{\SS}{\mathcal{S}}
\newcommand{\VV}{\mathcal{V}}
\newcommand{\WW}{\mathcal{W}}
\newcommand{\XX}{\mathcal{X}}

\newcommand{\B}[1]{\framebox{\rule{0pt}{.6em}\ensuremath{\!\tlachars #1\!}}\,}

\makeatletter \let\str=\@w \makeatother

\ifdraft
\long\def\ednote#1#2{\begin{quote}\framebox{\begin{minipage}{0.99\linewidth}\footnotesize\color{red} #1: #2\end{minipage}}\end{quote}}
\newcommand{\edmargin}[2]{\marginpar{\raggedright\footnotesize\color{red}#1: #2}}
\else
\long\def\ednote#1#2{}
\newcommand{\edmargin}[2]{}
\fi

\def\llnote{\ednote{LL}}

\def\smmargin{\edmargin{SM}}

\newtheorem{theorem}{Theorem}
\newtheorem{lemma}[theorem]{Lemma}
\newtheorem{definition}[theorem]{Definition}

\newenvironment{proofsketch}{\par\noindent\textbf{Proof (sketch).}\quad}{\medskip\par\noindent}

\let\tlachars\relax
\let\notla\relax
\newcommand{\deq}{\mathrel{\stackrel{\scriptscriptstyle\Delta}{=}}}
\def\A{\forall\,}
\def\E{\exists\,}
\newcommand{\TRUE}{\mbox{\sc true}}
\newcommand{\FALSE}{\mbox{\sc false}}
\newenvironment{noj}{\begin{array}[t]{@{}l@{}}}{\end{array}}

\makeatletter
\newcounter{abr@ctr}
\newcommand{\abr@c}{\c@abr@ctr\advance\c@abr@ctr\@ne}

\ifx\documentclass\undefined
\else
  \DeclareSymbolFont{tlaitalics}{\encodingdefault}{cmr}{m}{it}
  \let\itfam\symtlaitalics
\fi

\newcommand{\noTeXmath}{%
\c@abr@ctr=\itfam
\multiply\c@abr@ctr"100\relax
\advance\c@abr@ctr "7061\relax
\mathcode`a=\abr@c\mathcode`b=\abr@c\mathcode`c=\abr@c\mathcode`d=\abr@c
\mathcode`e=\abr@c\mathcode`f=\abr@c\mathcode`g=\abr@c\mathcode`h=\abr@c
\mathcode`i=\abr@c\mathcode`j=\abr@c\mathcode`k=\abr@c\mathcode`l=\abr@c
\mathcode`m=\abr@c\mathcode`n=\abr@c\mathcode`o=\abr@c\mathcode`p=\abr@c
\mathcode`q=\abr@c\mathcode`r=\abr@c\mathcode`s=\abr@c\mathcode`t=\abr@c
\mathcode`u=\abr@c\mathcode`v=\abr@c\mathcode`w=\abr@c\mathcode`x=\abr@c
\mathcode`y=\abr@c\mathcode`z=\abr@c
\c@abr@ctr=\itfam
\multiply\c@abr@ctr"100\relax
\advance\c@abr@ctr "7041\relax
\mathcode`A=\abr@c\mathcode`B=\abr@c\mathcode`C=\abr@c\mathcode`D=\abr@c
\mathcode`E=\abr@c\mathcode`F=\abr@c\mathcode`G=\abr@c\mathcode`H=\abr@c
\mathcode`I=\abr@c\mathcode`J=\abr@c\mathcode`K=\abr@c\mathcode`L=\abr@c
\mathcode`M=\abr@c\mathcode`N=\abr@c\mathcode`O=\abr@c\mathcode`P=\abr@c
\mathcode`Q=\abr@c\mathcode`R=\abr@c\mathcode`S=\abr@c\mathcode`T=\abr@c
\mathcode`U=\abr@c\mathcode`V=\abr@c\mathcode`W=\abr@c\mathcode`X=\abr@c
\mathcode`Y=\abr@c\mathcode`Z=\abr@c}

\makeatother
\noTeXmath

\begin{document}

\maketitle


\begin{abstract}
  We present
  a syntactic abstraction method to reason about
  first-order modal logics by using theorem provers for standard
  first-order logic and for propositional modal logic.


\end{abstract}


\section{Introduction}\label{sec:intro}


Verification of distributed and concurrent systems requires reasoning
about temporal behaviors.  A common approach is to express the
properties to be proved in a modal logic having one or more temporal
modalities.  For verifying real-world systems, a proof language must
%
%
also include equality, quantification, interpreted theories,
and local definitions.  It must
therefore encompass FOML (\emph{F}irst \emph{O}rder \emph{M}odal
\emph{L}ogic) and support operator definitions.  One such language is
\tlaplus \cite{lamport:tla+}, based on the logic TLA
that has two temporal modalities: the
usual $\Box$ (always) operator of linear-time temporal logic and 
$'$ (prime), a
%
%
restricted next-state operator such that $e'$ is the value of
$e$ at the next state if $e$ is an expression that does not contain a
modal operator.

A common way to prove an FOML sequent
$\Gamma\models\varphi$ ($\varphi$ holds in context $\Gamma$)
is to translate it to a semantically equivalent FOL sequent
$\Gamma^* \folmodels \varphi^*$
%
%
and to prove this FOL sequent.  For some FOMLs, this method is
semantically complete---that is, $\Gamma\models\varphi$ is valid iff
$\Gamma^* \folmodels \varphi^*$ is~\cite{ohlbach:translation}.
This approach has been followed for embedding FOML in
SPASS~\cite{hustadt:mspass}, Saturate~\cite{ganzinger:saturate}, and other theorem
provers.

Such a semantic translation may be appropriate for completely
automatic provers.  However, we are very far from being able to
automatically prove a formula that expresses a correctness property of
a non-trivial system.  A person must break the proof into smaller
steps that we call \emph{proof obligations}, usually by interacting
with the prover.  Requiring the user to interactively prove the
semantic translation of the FOML formula destroys the whole purpose of
using modal logic, which is to allow her to think in terms of the
simpler FOML abstraction of the theorem.  The user should therefore
decompose the FOML proof into FOML proof obligations.

In this paper we describe a method called \emph{coalescing} that handles many
FOML proof obligations by soundly abstracting them into formulas of either FOL
or propositional modal logic (ML). The resulting formulas are dealt with by
existing theorem provers for these logics. Although the basic idea of coalescing
is simple, some care has to be taken in the presence of equality and bound
variables. The translation becomes trickier in the presence of defined
operators.

\paragraph{Outline of this Paper.}

%
Section~\ref{sec:motivation} motivates our proposal by its application within
the \tlaplus Proof System TLAPS.
Section~\ref{sec:foml} formally introduces FOML and its two fragments,
FOL and ML\@.  Sections~\ref{sec:coalescing-modal} and
\ref{sec:coalescing-fol} present coalescing for modal and first-order
expressions respectively, proving their soundness.  Section
\ref{sec:leibnizing} extends the results to languages containing local
definitions.  In Section \ref{sec:safety} we outline a proof of the completeness of coalescing for proving safety properties. Section~\ref{sec:conclusion} discusses semantic translation vs.\
coalescing and suggests some optimizations and future work.

\section{Motivation} \label{sec:motivation}


\subsection{A Sample \tlaplus Proof}

\begin{figure}[bp]
  \centering
  \begin{minipage}[t]{.3\linewidth}
    \begin{tabular}[t]{@{}l@{}}
      $Init \deq \ldots$\\
      $Step \deq \ldots$\\
      $v \deq \ldots$\\
      $Spec \deq Init \land \Box[Step]_v$\\[.5ex]
      $Safe(p) \deq \ldots$
    \end{tabular}
  \end{minipage}
  \begin{minipage}[t]{.55\linewidth}
    \begin{tabular}[t]{@{}l@{}}
      \kw{theorem} $Spec \implies \A p \in Proc : \Box Safe(p)$\\
      \ps{1}{.} \kw{suffices}%
                  \begin{tabular}[t]{@{\ }l@{\ }l}
                    \kw{assume} & \kw{new} $p \in Proc$\\
                    \kw{prove}  & $Spec \implies \Box Safe(p)$
                  \end{tabular}\\[-.4em]
      \quad\kw{obvious}\\
      \ps{1}{1.} $Init \implies Safe(p)$\\
      \quad\kw{by} \kw{def} $Init$, $Safe$\\
      \ps{1}{2.}
         $Safe(p) \land [Step]_v \implies Safe(p)'$\\
      \quad\kw{by} \kw{def}
        $Safe$, $Step$, $v$\\
      \ps{1}{3.} \kw{qed}\\
      \quad\kw{by} \ps{1}{1}, \ps{1}{2},
      PTL \kw{def} $Spec$
    \end{tabular}
  \end{minipage}
  \caption{Proof of a safety property in TLAPS.}
  \label{fig:safety-example}
\end{figure}

Our motivation comes from designing the TLAPS proof
system~\cite{cousineau:tlaps} for \tlaplus, which can check
correctness proofs of complex, real-world algorithms~\cite{lamport:byzantizing}.
The essence of TLA proofs is to decompose proofs of temporal logic formulas
so that most of the obligations
contain no modal operator except \emph{prime}. Figure~\ref{fig:safety-example}
contains the outline of the proof of a simple safety property in TLAPS that
illustrates this decomposition. The system
specification is formula $Spec$, defined to equal
  $Init \land \Box[Step]_v$.
In this formula, $Init$ is a \emph{state predicate} that describes the possible
initial states, and $Step$ is an \emph{action predicate} that describes
possible state transitions. Syntactically: $Init$ is a FOL formula containing
state (a.k.a.\ flexible) variables; $Step$ is a formula containing state
variables, FOL operators, and the \emph{prime} operator; and
$v$ is a tuple of all state variables in the specification.
The formula $[Step]_v$ is a shorthand for $Step \lor (v'=v)$, and $\Box$ is the
usual ``always'' operator of temporal logic.
%
%
The temporal logic formula $Spec$ is evaluated over $\omega$-sequences of
states; it is true of a sequence $s_0 s_1 \ldots$ iff $Init$ is true at state
$s_0$ and, for all pairs of states $s_i$ and $s_{i+1}$, either $Step$ is true
or the value of $v$ does not change. The definitions of the formulas $Init$ and
$Step$, and the reason for writing $\Box[Step]_v$ instead
of $\Box Step$, are irrelevant in the context of this paper.
%
%
We wish to prove that a state formula $Safe(p)$ is true throughout any behavior
described by $Spec$, for every process $p \in Proc$.
%


The right-hand side of Figure~\ref{fig:safety-example} shows the
assertion and proof of the theorem.
The first step in the proof is purely first-order: it introduces a fresh
constant~$p$, assumes \mbox{$p \in Proc$}, and reduces the overall proof to showing the
implication $Spec \implies \Box Safe(p)$. Step $\ps{1}{1}$ asserts that the
initial condition implies $Safe(p)$. This formula does not contain any modal
operators.
%
%
Step $\ps{1}{2}$ shows that $Safe(p)$ is preserved by every transition
(as specified by $[Step]_v$).
The proof of this step is essentially first-order, although TLAPS must
handle the \emph{prime} modality.  The basic idea is to distribute
primes
inward in expressions using rules such as $(x+y)' = x'+y'$,
and then to replace the remaining primed expressions by new atoms.
For this example, we are
assuming that the specification is so simple that, after the
definitions of $Init$, $Next$, $v$, and $Safe$ have been expanded, the
FOL proof obligations generated for these two steps can be discharged
by a theorem prover.

Step $\ps{1}{3}$ concludes the proof.
It is justified by propositional temporal reasoning, in particular the principle
\[
  \begin{array}{@{}c}
    P \land A \implies P'\\
    \hline
    P \land \Box A \implies \Box P
  \end{array}
 \]
The \emph{PTL} in the step's proof tells TLAPS to invoke a PTL
decision procedure, which it does after replacing $Spec$ by its
definition and the formulas $Init$, $Safe(p)$ and $[Next]_v$ by fresh
atoms.  This effectively hides all operators other than those of
propositional logic, $\Box$, and \emph{prime}.

We call \emph{coalescing} the process of replacing expressions by atoms. It is
similar to the introduction of names for subformulas that theorem provers apply
during pre-processing steps such as CNF transformation. However, it has a
different purpose: the fresh names hide complex formulas that are meaningless to
a proof backend for a fragment of the original logic.
As explained in the example above, TLAPS uses coalescing in its translations to
invoke FOL and PTL backend provers, where the first do not support the modal
operators $\Box$ and \emph{prime}, and the second
do not support first-order constructs such as
quantification, equality or terms.
  \smmargin{Jul 3: added reference to KSAT}
An idea similar to coalescing underlies the KSAT decision
procedure~\cite{giunchiglia:ksat} for propositional multi-modal logic in that
modal top-level literals are abstracted by fresh atoms. Unlike KSAT, we consider
first-order modal logic, and we do not recursively construct formulas that must
be analyzed at deeper modal levels.

Coalescing cannot in itself be semantically complete because it cannot support
proof steps that rely on the interplay of the sublogics. For example,
separate FOL and PTL provers
cannot
prove rules that distribute quantifiers over temporal modalities. Similarly,
proofs of liveness properties via well-founded orderings essentially mix
quantification and temporal logic. However, we need very few such proof steps
in actual
proofs, and we can handle them using a more traditional
backend that relies on a FOL translation of temporal modalities. Coalescing is
complete for a class of temporal logic properties that includes safety
properties, which can be established by propositional temporal logic from
action-level hypotheses. For these applications, we have found coalescing to
be more flexible and more powerful in practice than a more traditional FOL
translation. In particular, proofs need not follow the simple schema of the
proof shown in Figure~\ref{fig:safety-example} but can invoke auxiliary
invariants or lemmas. The inductive reasoning underlying much of temporal logic
is embedded in PTL decision procedures but would be difficult to automate in a
FOL prover. On the other hand, the \emph{prime} modality by itself is simple
enough that it can be handled by a pre-processing step applied before passing
the proof obligation to a FOL prover.


\subsection{Coalescing In First-Order Modal Logic}

We believe that coalescing will be useful for proofs in
modal logics other than \tlaplus. We therefore present its fundamental
principles here using a simpler FOML containing a single modal operator $\Box$.
Corresponding to the translations we have implemented in TLAPS, we give two
translations of FOML obligations, one into FOL and the other into ML, and we
prove their soundness.

The idea underlying coalescing is very simple: abstract away a class
of operators by introducing a fresh atom in place of a subformula
whose principal operator is in that class.
However, doing this in a sound way in the presence of equality is not
trivial because of the \emph{Leibniz principle}, which asserts
 $(d=e) \Rightarrow (P(d) = P(e))$
for any expressions $d$ and $e$ and operator $P$.  The Leibniz
principle is valid in FOL but not FOML, which makes translating from
FOML obligations to FOL obligations tricky~\cite{mendelsohn:foml}.

For example, the formula $(v=0) \implies \Box(v=0)$ is not valid
in \tlaplus or more generally
in FOML when $v$ is flexible.
A naive application of standard FOL provers could
propagate the equality in the antecedent by substituting $0$ for $v$
throughout this formula, effectively applying the instance
$((v=0) = \TRUE) \implies (\Box(v=0) = \Box\TRUE)$
of the Leibniz principle, and consequently prove the formula using the axiom
$\Box\TRUE$. Such an approach is clearly unsound.
The standard translation of FOML into predicate logic~\cite{ohlbach:translation}
avoids this problem by making explicit the states at which formulas are
evaluated, but at the price of adding significant complexity to the
formula. Moreover, one
typically assumes specific properties about the accessibility relation(s)
underlying modal logics.  Incorporating these into first-order reasoning may
not be easy. For example, the $\Box$ modality of \tlaplus corresponds to the
transitive closure of the \emph{prime} modality, and this is not first-order
axiomatizable. Of course, whether this is an issue or not depends on the
particular modal logic one is interested in: semantic translation works very
well in applications such as~\cite{benzmueller:god} that are based on a modal
logic whose frame conditions are first-order axiomatizable.

Our approach is to coalesce expressions and formulas that are outside
the scope of a given theorem prover.  For the example above, coalescing
to FOL yields
\(  (v = 0) \implies \B{\Box(v=0)} \)
where $\B{\Box(v=0)}$ is a new $0$-ary predicate symbol, and this formula is
clearly not provable.
Similarly, coalescing to ML yields
\(  \B{v=0} \implies \Box\B{v=0}  \)
of propositional modal logic, and again, this formula is not provable.
We give a
detailed description of how to derive a new symbol $\B{exp}$ for an
arbitrary expression $exp$.  Care has to be taken when the coalesced expression
contains bound variables.
For example, a naive coalescing into FOL of the formula
  \mbox{$\A a : \Box(a=1) \implies a=1$},
which is valid over reflexive frames, would yield
  \mbox{$\A a : \B{\Box(a=1)} \implies a=1$},
from which we can deduce $\B{\Box(a=1)} \implies 0=1$ and then
  \mbox{$\A a: \lnot\Box(a=1)$},
which is clearly not valid. A correct coalescing yields
  \mbox{$\A a : \B{\Box(a=1)}(a) \implies a=1$}.

\paragraph{Operator Definitions.}

Coalescing is trickier for a language with
operator definitions like
%
  \,\,$P(x,y) \deq exp$\,,
%
where $exp$ does not contain free variables other than $x$ and $y$.
Definitions are necessary for structuring specifications and for
managing the complexity of proofs through lemmas about the defined
operators.  We therefore do not want to systematically expand all
defined operators in order to obtain formulas of basic FOML. The
Leibniz principle may not hold for an expression $P(a, b)$ if the
operator $P$ is defined in terms of modal operators---that is,
$(a=c) \land (b=d)$ need not imply
  $P(a,b)=P(c,d)$.  It would
therefore be unsound to encode $P$ as an uninterpreted predicate
symbol in first-order logic.  We show how soundness is preserved
by replacing an expression $P(a,b)$ with $\B{P,\eps_1,\eps_2}(a,b)$,
for some suitable expressions $\eps_1$ and $\eps_2$ (described 
in Section~\ref{sec:abstraction-defined} below),
where $\B{P,\eps_1,\eps_2}$ can be defined so it
satisfies the Leibniz
principle and also satisfies
\(
  \B{P,\eps_1,\eps_2}(a,b) = P(a,b)
\)
in suitably extended models of FOML, ensuring equisatisfiability of the original
and the coalesced formula.
Since it satisfies the Leibniz principle,
$\B{P,\eps_1,\eps_2}$ can be taken to be an uninterpreted
predicate symbol by a first-order theorem prover.
Our construction extends to the
case of definitions of second-order operators, which are allowed in \tlaplus.


%



\section{First-Order Modal Logic}
\label{sec:foml}
\subsection{Syntax.}

We introduce a language of first-order modal logic whose modal operator we
denote by $\modal$ in order to avoid confusion with the $\Box$ of \tlaplus.
The language omits the
customary distinction between function and predicate symbols, and
hence between terms and formulas.  This simplifies notation and allows our
results to apply to \tlaplus as well as to a conventional language
that does distinguish terms and formulas---the conventional
language just having a smaller set
of legal formulas.

We assume a first-order signature consisting of non-empty distinct denumerable sets
$\XX$ of rigid variables, $\VV$ of flexible variables, and $\OO$ of operator symbols.
Operator symbols have arities in $\nat$ and generalize both function and predicate symbols.
Expressions $e$ of FOML are then inductively defined by the following grammar:
 \[
  e\ \ ::=\ \
  x\ \,|\,\
  v\ \,|\,\
  op(e,\ldots,e)\ \,|\,\
  e = e\ \,|\,\
  \FALSE\ \,|\,\
  e \implies e\ \,|\,\
  \A x: e\ |\
  \modal e
 \]
%
%
where $x \in \XX$, $v \in \VV$, $op \in \OO$, and arities are
respected (empty parentheses are omitted for $0$-ary symbols).  We do
not allow quantification over flexible variables, so our flexible
variables are really ``flexible function symbols of arity 0''.
While \tlaplus\ allows quantification
over flexible variables,
it can be considered as another modal operator for the purposes of coalescing.

The notions of free and bound (rigid) variables are the usual ones. We say that
an expression is \emph{rigid} iff it contains neither flexible variables
nor subexpressions of the form
$\modal e$. The
standard propositional ($\TRUE$, $\lnot$, $\land$, $\lor$, $\equiv$)
and first-order ($\exists$) connectives are defined in the usual
way. The dual modality $\dual$ is introduced by defining $\dual e$ as
$\lnot\modal\lnot e$.
The extension to a multi-modal language is straightforward.

\subsection{Semantics.}

A \emph{Kripke model} $\MM$ for FOML
is a 6-tuple $(\II, \xi, \WW, R, \zeta, \modal_{\MM})$,
where:
\begin{itemize}
\item $\II$ is a standard first-order interpretation consisting of a universe
  $|\II|$ and, for every operator symbol $op$, an interpretation
  \(
    \II(op): |\II|^n \rightarrow |\II|
  \)
  where $n$ agrees with the arity of $op$. We assume that the universe $|\II|$
  contains two distinguished, distinct values $\true$ and $\false$.
\item $\xi: \XX \rightarrow |\II|$ is a valuation of the rigid variables.
\item $\WW$ is a non-empty set of states, and
  $R \subseteq \WW \times \WW$ is the accessibility relation.
\item $\zeta: \VV \times \WW \rightarrow |\II|$ is a valuation of the flexible
  variables at the different states of the model.
\item $\modal_{\MM}: 2^{|\II|} \fun |\II|$ is a function such that
  $\modal_{\MM}(S) = \true$ iff $S \subseteq \{\true\}$.
\end{itemize}
Note that we assume a constant universe, independent of the states of
the model, and we also assume that all operators in $\OO$
are rigid---i.e., interpreted independently of
the states.

We inductively define the interpretations of expressions $\notla\sem{e}^{\MM}_w$
at state $w$ of model $\MM$. When the model $\MM$ is understood from the context,
we drop it from the notation.
\begin{itemize}
\item $\notla\sem{x}^{\MM}_w \eqdef \xi(x)$\quad for $x \in \XX$
\item $\notla\sem{v}^{\MM}_w \eqdef \zeta(v,w)$\quad for $v \in \VV$
\item $\notla\sem{op(e_1, \ldots, e_n)}^{\MM}_w \eqdef \II(op)(\sem{e_1}^{\MM}_w, \ldots, \sem{e_n}^{\MM}_w)$\quad
  for $op \in \OO$
\item $\notla\sem{e_1 = e_2}^{\MM}_w \eqdef
  \left\{\begin{array}{l@{\ \ }l}
    \true & \mbox{if}\ \sem{e_1}^{\MM}_w = \sem{e_2}^{\MM}_w\\
    \false & \mbox{otherwise}
  \end{array}\right.$
\item $\notla\sem{\FALSE}^{\MM}_w \eqdef \false$
\item $\notla\sem{\varphi \implies \psi}^{\MM}_w \eqdef
  \left\{\begin{array}{l@{\ \ }l}
    \true & \mbox{if}\ \sem{\varphi}^{\MM}_w \neq \true\ \mbox{or}\ \sem{\psi}^{\MM}_w = \true\\
    \false & \mbox{otherwise}
  \end{array}\right.$
\item $\notla\sem{\A x: \varphi}^{\MM}_w \eqdef
  \left\{\begin{array}{l@{\ \ }l}
    \true & \mbox{if}\
            \begin{noj}
              \sem{\varphi}^{\MM'}_w = \true\
              \mbox{for all $\MM' =
              (\II,\xi',\WW,R,\zeta,\modal_{\MM})$ such that}\\
              \mbox{$\xi'(y) = \xi(y)$ for all $y \in \XX$ different from $x$}
            \end{noj}\\
    \false & \mbox{otherwise}
  \end{array}\right.$
\item $\notla\sem{\modal \varphi}^{\MM}_w \eqdef
  \modal_{\MM}(\{\sem{\varphi}^{\MM}_{w'} : (w,w') \in R\})$
\end{itemize}
We write $\MM,w \models \varphi$ instead of
$\notla\sem{\varphi}^{\MM}_w = \true$. We say that $\varphi$ is \emph{valid}
iff $\MM,w \models \varphi$ holds for
all $\MM$ and $w$, and that it is \emph{satisfiable} iff $\MM,w \models \varphi$ for
some $\MM$ and $w$.
We define a consequence relation
\,$\models$\, as follows (where $\Gamma$ is a set of formulas):
$\Gamma \models \varphi$ iff for all $\MM$, if $\MM, w \models \psi$
for all $\psi \in \Gamma$ and $w \in \WW$, then $\MM, w \models \varphi$ for
all $w \in \WW$.

Our definition of the semantics is a straightforward extension of the standard
Kripke semantics to our setting, where $\modal e$ need not denote a truth value.
The condition on the function $\modal_{\MM}$ used for interpreting $\modal$
ensures that $\MM,w \models \modal \varphi$ iff $\MM,w' \models \varphi$ for all
$w'$ such that $(w,w') \in R$ as in the standard Kripke semantics.
Because we assume a constant domain of interpretation, both Barcan formulas are
valid---that is, we have validity of%
\begin{equation}\label{eq:barcan}
  (\forall x : \modal \varphi)\ \equiv\ \modal(\forall x: \varphi).
\end{equation}
Moreover, since all operator symbols have rigid interpretations, it is easy to
prove by induction on the expression syntax that $\notla\sem{e}_w =
\sem{e}_{w'}$ holds for all states $w,w'$ whenever $e$ is a rigid expression.
It follows that implications of the form
\(
  \varphi \implies \modal \varphi
\)
are valid for rigid $\varphi$---for example:
\begin{equation}\label{eq:rigid-box}
  \A x,y : (x=y) \implies \modal(x=y).
\end{equation}


\subsection{FOL and ML fragments of FOML}

Two natural sublogics of FOML are first-order logic (FOL) and propositional modal
logic (ML).

FOL does not have flexible variables $\VV$ or expressions $\modal e$.
A first-order structure $(\II,\xi)$ consists of an interpretation $\II$ as above
and a valuation $\xi$ of the (rigid) variables.
The inductive definition of the
semantics consists of the relevant clauses of the one given above for FOML, and
the notions of first-order validity $\folmodels \,\varphi$, satisfiability, and
consequence carry over in the usual way.



ML does not have rigid variables, quantifiers, operator symbols or equality.
A (propositional) Kripke model for ML is given as $\KK = (\WW, R, \zeta)$ where
the set of states $\WW$ and the accessibility relation $R$ are as for FOML, and the
valuation $\zeta: \VV \times \WW \fun \{\true,\false\}$ assigns truth values to
flexible variables at every state. The inductive definition of
$\notla\sem{e}^{\KK}_w \in \{\true,\false\}$ specializes to the following clauses:

\begin{itemize}
\item \makebox[3cm][l]{$\sem{v}^{\KK}_w\ =\ \zeta(v,w)$} for $v \in \VV$
\item $\sem{\FALSE}^{\KK}_w\ =\ \false$
\item \makebox[3cm][l]{$\sem{\varphi \implies \psi}^{\KK}_w\ =\ \true$}
  iff\ \ $\sem{\varphi}^{\KK}_w = \false$ or $\sem{\psi}^{\KK}_w = \true$
\item \makebox[3cm][l]{$\sem{\modal \varphi}^{\KK}_w\ =\ \true$}
  iff\ \ $\sem{\varphi}^{\KK}_{w'} = \true$ for all $w' \in W$ such that $(w,w') \in R$
\end{itemize}
The notions of validity $\mlmodels\, \varphi$, satisfiability, and consequence
carry over as usual.

\section{Coalescing Modal Expressions}
\label{sec:coalescing-modal}

\subsection{Definition of the abstraction $\FOL{e}$}

One of our objectives is to apply standard first-order theorem provers
for proving theorems of FOML that are instances of first-order
reasoning.  Since the operator~$\modal$ is not available in first-order
logic, we must translate FOML formulas $\psi$ to purely
first-order formulas $\FOL{\psi}$ such that the consequence
$\FOL{\Gamma} \,\folmodels\, \FOL{\varphi}$ entails $\Gamma \models
\varphi$.  A naive but unsound approach would be to replace the modal
operator $\modal$ by a fresh monadic operator symbol $Nec$.
As explained in Section~\ref{sec:motivation}, this approach is not sound.
As we observed, a sound approach is to define $\FOL{\varphi}$ by using
the well-known standard translation from modal logic to first-order
logic~\cite{brauner:foml,ohlbach:translation} that makes explicit the
FOML semantics.  However, that translation introduces additional
complexity---complexity that is unnecessary for proof obligations that
follow from ordinary first-order reasoning.

Instead, we define $\FOL{\varphi}$ to be a syntactic first-order
abstraction of $\varphi$ in which modal subexpressions are
coalesced---that is, replaced by fresh operators.
If $\varphi$ is 
$(v=0) \implies \modal(v=0)$, then $\FOL{\varphi}$ is 
$(v = 0) \implies \B{\modal(v=0)}$,
where $\B{\modal(v=0)}$ is a new $0$-ary operator symbol.
%

We want to ensure that subexpressions appearing more than once are
abstracted by the same operators, allowing for instances of first-order theorems
to remain valid.  This requires some care for expressions that contain bound
variables. For example, we expect to prove
%
\begin{equation}\label{eq:ex-box}
  (\E x,z: \modal(v=x)) \equiv (\E y: \modal(v=y))
\end{equation}
We therefore define the fresh operator symbols
$\B{\modal e}$ as $\lambda$-abstractions over the
bound variables occurring in $e$, and these are identified modulo
$\alpha$-equivalence.  Formally, we let
 $\FOL{e} = \FOL{e^{\varepsilon}}$
where $\varepsilon$ denotes the empty list and,
for a list $\vec{y}$ of rigid variables, the first-order
expression $\FOL{e^{\vec{y}}}$ over the extended set of
variables $\XX \cup \VV$ is defined inductively as follows.
\begin{itemize}
\item $\FOL{x^{\vec{y}}} \eqdef x$ for $x \in \XX$ a rigid variable,
\item $\FOL{v^{\vec{y}}} \eqdef v$ for $v \in \VV$ a flexible variable,
\item $\FOL{(op(e_1, \ldots, e_n))^{\vec{y}}} \eqdef
  op(\FOL{(e_1)^{\vec{y}}}, \ldots, \FOL{(e_n)^{\vec{y}}})$
  for $op \in \OO$,
\item $\FOL{(e_1 = e_2)^{\vec{y}}} \eqdef \FOL{(e_1)^{\vec{y}}} = \FOL{(e_2)^{\vec{y}}}$,
\item $\FOL{\FALSE^{\vec{y}}} \eqdef \FALSE$
\item $\FOL{(e_1 \implies e_2)^{\vec{y}}} \eqdef \FOL{(e_1)^{\vec{y}}} \implies \FOL{(e_2)^{\vec{y}}}$,
\item $\FOL{(\A x : e)^{\vec{y}}} \eqdef \A x : \FOL{e^{x,\vec{y}}}$,
\item $\FOL{(\modal e)^{\vec{y}}} \eqdef \B{\lambda \vec{z}: \modal e}(\vec{z})$ where
  $\vec{z}$ is the subsequence of rigid variables in $\vec{y}$
   that appear free in $e$.
  (If $z$ is the empty sequence, this is simply $\B{\modal e}$.)
\end{itemize}
With these definitions, the formula (\ref{eq:ex-box}) is coalesced as
\begin{equation}\label{eq:ex-box-c}
  (\E x,z : \B{\lambda x: \modal(v=x)}(x)) \,\equiv\,
  (\E y: \B{\lambda y: \modal(v=y)}(y))
\end{equation}
which is an instance of the valid first-order equivalence%
  \[ (\E x,z: P(x)) \,\equiv\, (\E y: P(y)) \]
In particular, the two operator symbols occurring in (\ref{eq:ex-box-c}) are
identified because the two $\lambda$-expressions are $\alpha$-equivalent.
Identification of coalesced formulas modulo $\alpha$-equivalence ensures that
the translation is insensitive to the names of bound (rigid)
variables.  Section~\ref{sec:conclusion} discusses techniques for
abstracting from less superficial differences in first-order
expressions, such as between $\lambda x, y$ and $\lambda y, x$ and
between $a=b$ and $b=a$.

\subsection{Soundness of coalescing to FOL}

For a set $\Gamma$ of FOML formulas, we denote by $\FOL{\Gamma}$ the set of all
formulas $\FOL{\psi}$, for $\psi \in \Gamma$. We now show the soundness of the
abstraction.
\begin{theorem}\label{thm:coal-modal}
  For any set $\Gamma$ of FOML formulas and any FOML formula $\varphi$,
  if $\FOL{\Gamma} \,\folmodels \,\FOL{\varphi}$ then $\Gamma \models \varphi$.
\end{theorem}
\begin{proofsketch}
  Assume that $\Gamma \not\models \varphi$, so
  $\MM = (\II, \xi, \WW, R, \zeta, \modal_{\MM})$ is a Kripke model such that
  $\MM,w' \models \psi$ for all $\psi \in \Gamma$ and $w' \in \WW$, but that
  $\MM,w \not\models \varphi$ for some $w \in \WW$.

  For the extended set of variables $\XX \cup \VV$, define the first-order
  structure $\SS = (\II', \xi')$ where $\II'$ agrees with $\II$ for all operator
  symbols that appear in $\Gamma$ or $\varphi$, and where the valuation
  $\xi'$ is
  defined by $\xi'(x) = \xi(x)$ for $x \in \XX$
  and $\xi'(v) = \zeta(w,v)$ for $v
  \in \VV$. For the additional operator symbols introduced in $\FOL{\Gamma}$ and
  $\FOL{\varphi}$, we define
  \[
    \II'(\B{\lambda \vec{z}: \modal e})(d_1, \ldots, d_n)\ \ =\ \
    \sem{\modal e}^{\MM'}_w
  \]
  where $\MM'$ agrees with $\MM$ except for the valuation $\xi'$ that assigns
  the $i$\th variable of $\vec{z}$ to $d_i$. This interpretation is
  well-defined: if
  $\modal e_1$ and $\modal e_2$ are two expressions in $\Gamma$ or $\varphi$ that
  give rise to the same operator symbol, then $(\lambda \vec{z}_1 : \modal e_1)$
  and $(\lambda \vec{z}_2 : \modal e_2)$ must be $\alpha$-equivalent, and
  therefore $\II'(\B{\lambda \vec{z}_1: \modal e_1})(d_1, \ldots, d_n) =
  \II'(\B{\lambda \vec{z}_2: \modal e_2})(d_1, \ldots, d_n)$.

  It is straightforward to prove that $\notla\sem{\FOL{e}}^{\SS} = \sem{e}^{\MM}_w$
  holds for all expressions $\FOL{e}$ that appear in $\FOL{\Gamma}$ or
  $\FOL{\varphi}$. In particular, it follows that $\SS \,\folmodels \,\FOL{\psi}$
  for all $\psi \in \Gamma$ and $\SS \,\nfolmodels\, \FOL{\varphi}$. This shows that
  $\FOL{\Gamma} \,\nfolmodels\, \FOL{\varphi}$ and concludes the proof.
  \qed
\end{proofsketch}

\section{Coalescing First-Order Expressions}
\label{sec:coalescing-fol}

We now define an abstraction $\ML{\varphi}$ of FOML formulas to formulas of
propositional modal logic. Again, we require for soundness that $\Gamma \models
\varphi$ whenever $\ML{\Gamma}\, \mlmodels \,\ML{\varphi}$---that is, consequence
between abstracted formulas implies consequence between the original ones. In
this way, we can use theorem provers for propositional modal logic to carry out
FOML proofs that are instances of propositional modal reasoning. The abstraction
$\ML{\varphi}$ replaces all first-order subexpressions $e$ of $\varphi$ by new
(propositional) flexible variables $\B{e}$, where variables $\B{\A x:e}$
are
once again identified modulo $\alpha$-equivalence. Formally, the translation is defined
as follows.
\begin{itemize}
\item $\ML{x} \eqdef \B{x}$ for $x \in \XX$ a rigid variable,
\item $\ML{v} \eqdef v$ for $v \in \VV$ a flexible variable,
\item $\ML{(op(t_1, \ldots, t_n))} \eqdef \B{op(t_1, \ldots, t_n)}$
    for $op\in \OO$,
\item $\ML{(e_1 = e_2)} \eqdef \B{e_1 = e_2}$,
\item $\ML{\FALSE} \eqdef \FALSE$,
\item $\ML{(e_1 \implies e_2)} \eqdef \ML{(e_1)} \implies \ML{(e_2)}$,
\item $\ML{(\A x: e)} \eqdef \B{\A x: e}$,
\item $\ML{(\modal e)} \eqdef \modal \ML{e}$.
\end{itemize}
As an example,
coalescing the formula
%
  \[ (x=y) \,\land\, \modal\dual\TRUE \,\implies\, \modal\dual(x=y) \]
%
yields the ML-formula
\begin{equation}\label{eq:box-eq-c}
  \B{x=y} \,\land\, \modal\dual\TRUE \,\implies\, \modal\dual\B{x=y}
\end{equation}
The implication (\ref{eq:box-eq-c}) is not ML-valid. However,
for rigid variables $x$ and $y$,
it follows from
the hypothesis $\B{x=y} \implies \modal\B{x=y}$, which is justified by the FOML
law (\ref{eq:rigid-box}).

For a set $\Gamma$ of FOML formulas, we denote by $\ML{\Gamma}$ the set of modal
abstractions $\ML{\psi}$, for all $\psi \in \Gamma$. Moreover, we define the set
$\HH(\Gamma)$ to consist of all formulas of the form $\B{e} \implies
\modal\B{e}$, for all flexible variables $\B{e}$ introduced in $\ML{\Gamma}$ that
correspond to rigid expressions $e$ in $\Gamma$.

\begin{theorem}
  Assume that $\Gamma$ is a set of FOML formulas and that $\varphi$ is a FOML formula.
  If\/ $\ML{\Gamma}, \HH(\Gamma \cup \{\varphi\}) \,\mlmodels\, \ML{\varphi}$
  then $\Gamma \models \varphi$.
\end{theorem}
\begin{proofsketch}
  As in Theorem~\ref{thm:coal-modal}, we prove the contra-positive.
  Assume that $\MM = (\II,\xi,\WW,R,\zeta,\modal_{\MM})$ is a Kripke model such that $\MM,w'
  \models \psi$ for all $\psi \in \Gamma$ and $w' \in \WW$, but $\MM,w
  \not\models \varphi$ for a certain $w \in \WW$.

  Define the propositional Kripke model $\KK = (\WW,R,\zeta')$ where $\zeta'$
  assigns truth values in $\{\true,\false\}$ to all states $w' \in \WW$ and
  flexible variables in $\ML{\Gamma}$ or $\ML{\varphi}$:
  \[\begin{array}{@{}l}
    \zeta'(w',v) = \true\ \ \mbox{iff}\ \ \zeta(w',v) = \true\ \ \mbox{for $v \in \VV$}\VS\\
    \zeta'(w',\B{x}) = \true\ \ \mbox{iff}\ \ \xi(x) = \true\ \ \mbox{for $x \in \XX$}\VS\\
    \zeta'(w',\B{op(t_1,\ldots,t_n)}) = \true\ \ \mbox{iff}\ \
      \notla\sem{op(t_1,\ldots,t_n)}^{\MM}_{w'} = \true\VS\\
    \zeta'(w',\B{e_1 = e_2}) = \true\ \ \mbox{iff}\ \
      \notla\sem{e_1}^{\MM}_{w'} = \sem{e_2}^{\MM}_{w'}\VS\\
    \zeta'(w',\B{\A x:e}) = \true\ \ \mbox{iff}\ \ \MM,w' \models \A x:e
  \end{array}\]
  Again,
  $\zeta'$ is well-defined. It is easy to prove, for all $w' \in \WW$
  and all $e$ such that $\ML{e}$ appears in $\ML{\Gamma}$ or
  $\ML{\varphi}$, that
  $\KK,w' \models \ML{e}$ iff $\notla\sem{e}^{\MM}_{w'} = \true$.
  In particular, it follows that $\KK,w' \models \ML{\psi}$ for
  all $\psi \in \Gamma$ and that $\KK,w \,\nmlmodels\, \ML{\varphi}$.

  Furthermore, the definition of $\KK$ ensures that $\KK,w' \models \psi$ holds for
  all $\psi \in \HH(\Gamma \cup \{\varphi\})$ and all $w' \in \WW$
  because $\notla\sem{e}^{\MM}_{w'} =
  \sem{e}^{\MM}_{w''}$ holds for all rigid expressions $e$
  and all states $w',w'' \in \WW$.

  It follows that $\ML{\Gamma}, \HH(\Gamma \cup \{\varphi\})
  \,\nmlmodels\, \ML{\varphi}$, which concludes the proof.~\qed
%
\end{proofsketch}

\section{Coalescing in the presence of operator definitions}
\label{sec:leibnizing}

\subsection{Operator definitions}
\label{sec:definitions}

We now extend our language to allow definitions of the form
\[
  d(x_1,\ldots,x_n)\ \deq\ e
\]
where $d$ is a fresh symbol,
$x_1,\ldots,x_n$ are pairwise distinct rigid
variables, and $e$ is an expression whose free rigid variables are among
$x_1,\ldots,x_n$.


For an operator $d$ defined as above and expressions $e_1,\ldots,e_n$, the
application $d(e_1,\ldots,e_n)$ is a well-formed expression whose semantics is
given by:
\[\notla
  \sem{d(e_1,\ldots,e_n)}^{\MM}_w =
  \sem{e[e_1/x_1, \ldots, e_n/x_n]}^{\MM}_w
\] In other words, the defining expression is evaluated when the
arguments have been substituted for the variables.  However, when
reasoning about expressions containing defined operators, one does not
wish to systematically expand definitions.  If the precise definition
is unimportant, it is better to leave the operator unexpanded in order
to keep the formulas small.  We now extend the coalescing techniques
introduced in the preceding sections to handle expressions that may
contain defined operators.

It is easy
to see that the algorithm introduced in
Section~\ref{sec:coalescing-fol} for abstracting first-order subexpressions
remains sound if we handle defined operators like operators in $\OO$.
In particular, two expressions $d(\vec{e}_1)$ and $d(\vec{e}_2)$
are abstracted by the same flexible variable only if they are syntactically
equal up to $\alpha$-equivalence.
However, this simple approach does not work for
the algorithm of Section~\ref{sec:coalescing-modal} that abstracts
modal subexpressions. As an example, consider the
definition
\begin{equation}\label{eq:def-cst}
  cst(x)\ \deq\ \E y: \modal(x=y)
\end{equation}
and the formula
\begin{equation}\label{eq:use-cst}
  (u=v) \,\implies\, (cst(u) \equiv cst(v))
\end{equation}
where $u$ and $v$ are flexible variables.  An expression $e$ satisfies
$cst(e)$ at state $w$ iff the value of $e$ is the same at all
reachable states $w'$.  Hence, formula (\ref{eq:use-cst}) is obviously
not valid.  If $cst$ were treated like an operator in $\OO$, the
algorithm of Section~\ref{sec:coalescing-modal} would leave
(\ref{eq:use-cst}) unchanged.  However, $u$ and $v$ would be
considered ordinary (rigid) variables and $cst$ would be considered an
uninterpreted operator symbol, so (\ref{eq:use-cst}), seen as a FOL
formula, would be
provable.  Thus, it would be unsound to simply treat defined operators
like operators in $\OO$ in our algorithm for coalescing modal
subexpressions.

\subsection{Rigid arguments and Leibniz positions}
\label{sec:leibniz-cases}

The example above shows that in the presence of definitions, FOML formulas
without any
visible modal operators may violate the Leibniz principle that
substituting equals for equals should yield equal results. However, a first
observation shows that the Leibniz principle still holds for rigid arguments.

\begin{lemma}\label{thm:rigid-leibniz}
  For any defined $n$-ary operator $d$, expressions $e_1, \ldots, e_n$, any $i
  \in 1\,..\,n$ with
  $e_i$ rigid, Kripke model $\MM$, state $w$, and
  rigid variable $x$ that does not occur free in any $e_j$, we have
  \[\notla
    \sem{d(e_1,\ldots,e_n)}^{\MM}_w =
    \sem{d(e_1,\ldots,e_{i-1},x,e_{i+1},\ldots,e_n)}^{\MM'}_w
  \]
  where $\MM'$ agrees with $\MM$ except for the valuation $\xi'$ of rigid
  variables, which is like $\xi$ but assigns $x$ to $\notla\sem{e_i}^{\MM}_w$.
\end{lemma}
\begin{proofsketch}
  Since $e_i$ is rigid, the value of
  $\notla\sem{e_i}^{\MM}_{w'}$, for any $w' \in W$, is independent of the state
  $w'$. The assertion is then proved by induction on the defining expression for
  operator $d$.
  \qed
\end{proofsketch}
For a non-rigid argument of a defined operator, the Leibniz principle is
preserved when the argument does not appear in a modal context in the defining
expression. We inductively define which argument positions of an FOML operator or
connective are Leibniz (satisfy the Leibniz principle).

\pagebreak 

\begin{definition}[Leibniz argument positions]\label{def:leibniz-pos}\mbox{}
  \begin{itemize}
  \item All argument positions of the operators in $\OO$ and of all FOML
    connectives except $\modal$ are Leibniz. The single argument position of $\modal$
    is not Leibniz.
  \item For an operator defined by $d(x_1,\ldots,x_n) \deq e$, the
    $i$\th argument position of $d$ is Leibniz iff
    $x_i$ does not occur within a non-Leibniz argument position in $e$.
  \end{itemize}
\end{definition}
In other words, the $i$\th argument position of a defined operator is Leibniz iff
the $i$\th parameter does not appear in the scope of any
occurrence of $\modal$
in the full expansion of the defining expression.

\begin{lemma}\label{thm:leibniz-pos}
  Assume that $d$ is a defined $n$-ary operator whose $i$\th argument position
  is Leibniz (for $i \in 1\,..\,n$). For any expressions $e_1, \ldots, e_n$, Kripke model
  $\MM$, state $w$ and rigid variable $x$ that does not occur free in any $e_j$,
  we have
  \[\notla
    \sem{d(e_1,\ldots,e_n)}^{\MM}_w =
    \sem{d(e_1,\ldots,e_{i-1},x,e_{i+1},\ldots,e_n)}^{\MM'}_w
  \]
  where $\MM'$ agrees with $\MM$ except for the valuation $\xi'$ of rigid
  variables, which is like $\xi$ but assigns $x$ to $\notla\sem{e_i}^{\MM}_w$.
\end{lemma}
\begin{proofsketch}
  Induction on the syntax of the defining expression for $d$.~\qed
\end{proofsketch}%
It follows from Lemmas~\ref{thm:rigid-leibniz} and~\ref{thm:leibniz-pos} that
the implication
\[\begin{noj}
  (e_{i}=f) \;\implies\; 
  (d(e_1,\ldots,e_{n}) = d(e_1,\ldots,e_{i-1},f,e_{i+1},\ldots,e_{n}))
\end{noj}\]
is valid when $e_{i}$ and $f$ are rigid expressions or when the
$i$\th argument position of $d$ is Leibniz.

\subsection{Coalescing for defined operators}
\label{sec:abstraction-defined}

The definition
of the syntactic abstraction $\FOL{e}$ for the extended language
is now completed by defining
\begin{itemize}
\item $\FOL{(d(e_1, \ldots, e_n))^{\vec{y}}} \eqdef
  \B{d,\epsilon_1,\ldots,\epsilon_n}(\FOL{(e_1)^{\vec{y}}},\ldots,\FOL{(e_n)^{\vec{y}}})$
  for a defined $n$-ary operator $d$ where
  \[\begin{array}{@{}ll}
    \epsilon_i = \ast &
    \mbox{if the $i$\th position of $d$ is Leibniz or $e_i$ is
      a rigid expression,}\\
    \epsilon_i = e_i & \mbox{otherwise.}
  \end{array}\]
\end{itemize}
With these definitions,
the single argument position of operator $cst$
introduced by (\ref{eq:def-cst}) is not Leibniz, and
coalescing formula (\ref{eq:use-cst}) yields
\[
  (v=w) \,\implies\, (\B{cst,v}(v) \equiv \B{cst,w}(w))
\]
for two distinct fresh operators $\B{cst,v}$ and $\B{cst,w}$. As expected, this
formula cannot be proved. However, the formula
\(
  \A x,y: (x=y) \,\implies\, (cst(x) \equiv cst(y))
\)
is coalesced as
\(
  \A x,y: (x=y) \,\implies\, (\B{cst,\ast}(x) \equiv \B{cst,\ast}(y))
\)
and is valid.

\pagebreak 

\begin{theorem}\label{thm:coal-def}
  Theorem~\ref{thm:coal-modal} remains valid for FOML formulas in the presence
  of defined operator symbols.
\end{theorem}
\begin{proofsketch}
  Extending the proof of Theorem~\ref{thm:coal-modal}, we define the
  interpretation of the fresh operator symbols as follows:
  \[\notla\begin{noj}
    \II'(\B{d,\epsilon_1,\ldots,\epsilon_n})(d_1,\ldots,d_n)\ =\
    \sem{d(\alpha_1,\ldots,\alpha_n)}^{\MM'}_w\\[2mm]
    \mbox{where}\ \alpha_i =
    \left\{\begin{array}{l@{\ \ }l}
        e_i & \mbox{if } \epsilon_i = e_i\\
        x_i & \mbox{if } \epsilon_i = \ast
    \end{array}\right.
  \end{noj}\]
  In this definition, $w$ is the state fixed in the proof and $\MM'$ agrees with
  $\MM$ except for the valuation $\xi'$ that assigns the variables $x_i$ to $d_i$.

  Again, one proves that $\notla\sem{\FOL{e}}^{\SS} = \sem{e}^{\MM}_w$ for all
  expressions $\FOL{e}$ that appear in $\FOL{\Gamma}$ or $\FOL{\varphi}$. For
  the expressions corresponding to applications of defined operators, the
  proof is obvious for those arguments where $\epsilon_i = e_i$, and it
  makes use
  of Lemmas~\ref{thm:rigid-leibniz} and~\ref{thm:leibniz-pos} when $\epsilon_i =
  \ast$.~\qed
\end{proofsketch}

\section{Proving Safety Properties by Coalescing in TLA}
\label{sec:safety}

We now give evidence for the usefulness of coalescing by showing that it can be
the basis for a complete proof system for establishing safety properties in
\tlaplus, assuming the ability to prove any valid first-order formula in the
set-theoretic language underlying \tlaplus. In fact, we argue that proofs of
arbitrary safety properties can be transformed into the form used in
Section~\ref{sec:motivation}.

In \tlaplus, properties of systems are expressed as temporal logic formulas,
which are evaluated over infinite sequences of states. We say that a formula is
true of a finite sequence of states if it is true for some infinite extension of
that finite sequence. A formula expresses a \emph{safety property} if it holds
for an infinite sequence of states if and only if it holds for every finite
prefix of that sequence.

The standard form of a \tlaplus specification is $Init \land \Box[Next]_v \land
L$ where $Init$ is a state predicate (a first-order formula that contains only
unprimed state variables), $v$ is a tuple containing all state variables, $Next$
  \smmargin{Jul 3: slight rewording}
is an action formula (a formula of first-order logic extended by the \emph{prime}
operator), and $L$ is a conjunction of fairness conditions that are irrelevant
for proving safety properties. In order to prove that the specification
establishes a safety property $P$, we need to establish the theorem
\[
  Init \land \Box[Next]_v \implies P.
\]
The first step is to reformulate the problem as an \emph{invariant assertion} of
the form
\begin{equation}\label{eq:tla-inv}
  HInit \land \Box[HNext]_{hv} \implies \Box Inv
\end{equation}
where $Inv$ is a state predicate. A general way to do this is to add a
\emph{history variable}~\cite{abadi:existence} to the original specification.
The history variable records the sequence of states seen so far, and $Inv$ is
true for a given value of the history variable if and only if the safety
property $P$ is true for the corresponding sequence of states.

\pagebreak 

The second step is to prove the invariance of $Inv$ 
by using an \emph{inductive
  invariant}---a state predicate $IInv$ such that all of the following formulas
are valid:
\begin{enumerate}
\item $HInit \implies IInv$
\item $IInv \land [HNext]_{hv} \implies IInv'$
\item $IInv \implies Inv$
\end{enumerate}

Assuming that formula \eqref{eq:tla-inv} is valid, a suitable inductive
invariant $IInv$ exists under standard assumptions on the expressiveness of the
language of state predicates (see, e.g., \cite{Apt:1981}), and these assumptions
are satisfied by the set-theoretic language underlying \tlaplus.

By coalescing to propositional temporal logic using the techniques described in
Section~\ref{sec:coalescing-fol}, a PTL decision procedure easily checks that
\eqref{eq:tla-inv} follows from facts (1)--(3) above. In practice, we find
it preferable to verify inferences in temporal logic by appealing to a decision
procedure rather than by applying a fixed set of rules because the
user is then free
to structure the proof in the most convenient way. For example, the inductive
invariant could be split into several mutually inductive formulas, for which the
corresponding facts may be easier to check than (1)--(3) in the
standard invariance proof above. On the other hand, the inductive nature of the
typical temporal logic proofs such as the one above makes it unrealistic to
expect that an ordinary FOL prover would be able to establish the FOL
translation of \eqref{eq:tla-inv} from (1)--(3).

Completing the proof of \eqref{eq:tla-inv} requires proving 
formulas (1)--(3). Observe that (1) and (3) are state predicates
and thus ordinary first-order formulas in \tlaplus's set theory. By assumption,
their proofs can be discharged by the underlying FOL prover.

Formula (2) is an action formula and therefore contains \tlaplus's \emph{prime}
operator in addition to first-order logic, but no other operator of temporal
logic. We now show how we reduce the proof of an action formula $A$ to FOL
proofs with the help of coalescing.

We begin by eliminating all defined operators that occur in $A$, expanding their
definitions. This results in an action formula $B$ that is equivalent to $A$,
but that only contains built-in \tlaplus operators. Second, we use the fact that
\emph{prime} distributes over all non-modal built-in \tlaplus operators and rewrite
formula $B$ to an equivalent action formula $C$ in which \emph{prime} is only
applied to flexible variables.\footnote{%
  The syntax of \tlaplus ensures that \emph{prime} cannot be nested.
} %
Finally, we apply the coalescing technique described in
Section~\ref{sec:coalescing-modal} for abstracting the \emph{prime} modality,
and obtain a formula $\FOL{C}$. It 
can be shown
that for the restricted
fragment where all modal expressions are of the form $v'$ for flexible
variables $v$, the formula $C$ is FOML-valid if and only if $\FOL{C}$ is FOL-valid,
and therefore the
underlying FOL prover will be able to prove formula (2) above.

The procedure above shows how in theory the two forms of coalescing
that we have proposed in this paper can be complete for proving safety
properties in \tlaplus, assuming that we have a complete proof
procedure for the set theory underlying \tlaplus.  In practice, no
such procedure can exist and it is preferable to keep formulas small,
so it is important not to expand all definitions.  These practical
considerations underlie the handling of defined operators described in
Section~\ref{sec:leibnizing}.  
 \llnote{2 July: I rewrote the rest of this paragraph on liveness,
  saying a tiny bit more about \tlaplus\ liveness proofs,
  tightening the prose, and removing text that said steps 
  requiring a full FOML backend prover are ``rare''.  
  Note that this material is largely repeated in the 2nd paragraph
  of the conclusion.\\[.5em]
  SM: Leslie, thanks for the rewrite. I was aware of the redundancy with the
  conclusion but considered some repetition to be acceptable, given that
  IJCAR reviewers apparently didn't get this point.
}
Although we have focused on safety properties, for which we can obtain
a completeness result, the same overall technique can also be applied
to the proof of liveness properties.  In TLA, liveness properties are
also proved by combining action-level reasoning and temporal logic proof
rules.  Most of the steps in such a proof can be reduced by coalescing
to FOL or propositional ML reasoning.  However, there will often be a
few steps that require non-propositional temporal logic
reasoning---for example, because they are based on well-founded
orderings.  Because there are not many such steps, a backend prover
for them with a low degree of automation should be acceptable.

\section{Conclusion}
\label{sec:conclusion}

We presented a technique of coalescing that allows a user to decompose the proof
of FOML formulas into purely first-order and purely propositional modal
reasoning. This technique is inspired by reasoning about \tlaplus specifications
and has been implemented in the \tlaplus Proof System, where we have found it
useful for verifying temporal logic properties.
In particular, the overwhelming majority of proof
obligations that arise during \tlaplus proofs
contain only the \emph{prime} modal operator. For this fragment,
rewriting by the valid equality $op(e_1,\ldots,e_n)' = op(e_1',\ldots,e_n')$, for
operators $op \in \OO$, followed by coalescing to FOL is complete. Many of the
proof obligations that involve the $\Box$ modality of \tlaplus are instances of
propositional temporal reasoning, and these can be handled by coalescing to
ML and invoking a decision procedure for propositional temporal logic.

Coalescing to FOL eschews semantic translation of FOML
formulas~\cite{ohlbach:translation} in favor of replacing a subformula whose
principal operator is modal by a fresh operator symbol.
The resulting formulas
are simpler than those obtained by semantic translation, and they can readily be
understood in terms of the original FOML formulation of the problem. Coalescing
is not a complete proof procedure in itself. For example, the valid Barcan
formula (\ref{eq:barcan})
cannot be proved using only our two translations.  TLA proofs contain
only a small number of such proof obligations, and we expect TLAPS to
be able to handle them with a semantic translation to FOL\@.
In our context, the validity problem of first-order temporal logic is
$\Pi_1^1$-complete, so incompleteness should not be considered an
argument against the use of coalescing. Semantic translation of temporal logic
would require inductive reasoning over natural numbers, and we could not even
expect simple proofs such as the one shown in Section~\ref{sec:motivation} to be
discharged automatically, whereas proof by coalescing benefits from efficient
decision procedures for propositional temporal logic.
For applications other than \tlaplus theorem proving that require
first-order modal reasoning, the trade-off in choosing between semantic
translation and coalescing will depend upon how effective one expects
semantic translation and standard first-order theorem proving to work
in practice.  One recent experiment~\cite{benzmueller:god} found this
technique entirely satisfactory, but it used a modal logic too weak to
handle the applications that concern us.

The definition of coalescing to FOL presented in
Section~\ref{sec:coalescing-modal} identifies modal subformulas such as
(\ref{eq:ex-box}) that are identical up to the names of bound rigid variables
that they contain. This definition can be refined to identify formulas that
differ in less superficial ways.
For example, it may be desirable to reorder bound variables according to their
appearance in coalesced subformulas. This would allow us to coalesce the formula
\[
  (\E y\, \A x : \Box P(x,y)) \implies (\A x\, \E y : \Box P(x,y))
\]
to the valid FOL formula
\[
  (\E y\, \A x : \B{\lambda x,y : \Box P(x,y)}(x,y)) \implies
  (\A x\, \E y : \B{\lambda x,y : \Box P(x,y)}(x,y))
\]
rather than the formula
\[
  (\E y\, \A x : \B{\lambda y,x : \Box P(x,y)}(y,x)) \implies
  (\A x\, \E y : \B{\lambda x,y : \Box P(x,y)}(x,y))
\] obtained according to the definition given in
Section~\ref{sec:coalescing-modal}, which results in the two fresh
operators being distinct.  In general, we would like coalesced
versions of different expressions to use the same atomic symbol
wherever that would be valid.  For example,
$\B{e_{1}=e_{2}}$ and $\B{e_{2}=e_{1}}$ could be the same symbol.
%
%

Rewriting a formula before coalescing can also make the translated
obligation easier to prove.  For example, the formula $\Box e$ for a
rigid expression $e$ can be replaced by $\B{\Box\FALSE} \lor e$.
In a modal logic whose $\Box$ modality is
reflexive, the disjunct $\B{\Box\FALSE}$ is not necessary.
In this way, the formula
\[
  \A x,y : \Box(x=y) \implies \Box(f(x) = f(y))
 \]
for $f \in \OO$ could be proved directly by
translating with coalescing to FOL instead of requiring two steps, the
first proving $(x=y) \implies (f(x)=f(y))$ with FOL and the second being
translated to ML\@.
Another such rewriting is distributing TLA's modal \emph{prime} operator
over rigid operators used by TLAPS when translating to FOL\@.
%

We don't know yet if optimizations of the translations beyond those we
have already implemented in TLAPS will be useful in practice.  So far,
we have proved only safety properties for realistic algorithms, which
in TLA requires little temporal reasoning.  We have begun writing
formal liveness proofs, but TLAPS will not completely check them until
we have a translation that can handle formulas which, like the Barcan
formula, inextricably mix quantifiers and modal operators.

%
\makeatletter
\def\realslash{/}
\begingroup
\catcode`\/\active
\catcode`\.\active
\catcode`:\active
\gdef\urlslash{\@ifnextchar/{\doubleslash}{\discretionary{}{}{}\realslash}}
\gdef\urlend#1{\let/\urlslash\let.\urldot
                 \discretionary{}{}{}#1\discretionary{}{}{}\endgroup}
\endgroup
\def\urldot{.\discretionary{}{}{}}
\def\url{\begingroup\urlbegin}
\def\urlbegin{
                       \catcode`\~12\relax
                       \catcode`\#12\relax
                       \catcode`\$12\relax
                       \catcode`\&12\relax
                       \catcode`\_12\relax
                       \catcode`\^12\relax
                       \catcode`\\12\relax
                       \catcode`\/\active
                       \catcode`\.\active
                       \tt
                       \urlend}
\def\doubleslash#1{\discretionary{}{}{}//}

\makeatother

\bibliographystyle{plain}
\bibliography{bib}
\end{document}